\documentclass[conference]{IEEEtran}
\IEEEoverridecommandlockouts
\usepackage{fancyhdr}

\usepackage{cite}
\usepackage{amsmath,amssymb,amsfonts}
\usepackage{algorithmic}
\usepackage{graphicx}
\usepackage{textcomp}
\usepackage{xcolor}
\usepackage{xspace}
\usepackage{tabularx}
\usepackage{booktabs}
\usepackage{listings}
\usepackage{threeparttable}
\usepackage{url}
\renewcommand{\thetable}{\Roman{table}}
\def\BibTeX{{\rm B\kern-.05em{\sc i\kern-.025em b}\kern-.08em
    T\kern-.1667em\lower.7ex\hbox{E}\kern-.125emX}}

\newcommand{\chatrep}{\textsc{ChatRepair}\xspace}
\newcommand{\chatreponeitersh}{\textsc{OneIter-SH}\xspace} 
\newcommand{\chatreponeiterm}{\textsc{OneIter-M}\xspace}
\newcommand{\chatgpt}{ChatGPT\xspace}

\newcommand{\conclusion}[1]{%
  \begin{center}
    \fcolorbox{black}{gray!8}{\parbox{.9\linewidth}{\textit{\textbf{#1}}}}
  \end{center} 
}

\begin{document}

\title{Studying and Understanding the Effectiveness and Failures 
of Conversational LLM-Based Repair}


\author{
    \IEEEauthorblockN{Aolin Chen$^a$$^c$,
    Haojun Wu$^a$$^c$,
    Qi Xin$^a$$^c$$^{\ast}$,\thanks{\vspace{-4pt}$^{\ast}$ Corresponding author}}
    Steven P. Reiss$^b$,
    Jifeng Xuan$^a$
    \IEEEauthorblockA{$^a$ School of Computer Science, Wuhan University, China}
    \IEEEauthorblockA{$^b$ Department of Computer Science, Brown University, USA}
    \IEEEauthorblockA{$^c$ Hubei Luojia Laboratory, China}
    \IEEEauthorblockA{\{aolin.chen, haojunwu, qxin\}@whu.edu.cn, spr@cs.brown.edu, jxuan@whu.edu.cn}}

\maketitle

\begin{abstract}

Automated program repair (APR) is designed to automate the process of bug-fixing.
In recent years, thanks to the rapid development of large language models (LLMs),
automated repair has achieved remarkable progress.
Advanced APR techniques powered by conversational LLMs,
most notably ChatGPT, have exhibited impressive repair abilities
and gained increasing popularity due to the capabilities of the underlying LLMs
in providing repair feedback and performing iterative patch improvement.
Despite the superiority, conversational APR techniques
still fail to repair a large number of bugs.
For example, a state-of-the-art conversational technique
\chatrep does not correctly repair over half of the single-function bugs in the Defects4J dataset.
To understand the effectiveness and failures of conversational LLM-based repair
and provide possible directions for improvement,
we studied the exemplary \chatrep with a focus on 
comparing the effectiveness of its cloze-style and full-function repair strategies,
assessing its key iterative component for patch improvement,
and analyzing the repair failures. 
Our study has led to a series of findings, 
which we believe provide key implications for future research.
\end{abstract}


\section{Introduction}

Automated program repair (APR) aims to alleviate a developer's burden
by automatically identifying buggy code and proposing and validating patches for it.
Powered by the large language models (LLMs),
advanced APR techniques have demonstrated remarkable repair abilities.
Among them, conversational APR techniques,
which use a conversational LLM (most notably \chatgpt)
to understand the failure and generate patches,
have gained increasing attention due to 
the conversational LLM's unique capabilities
in providing repair accompanying feedback,
which helps the developer understand the repair solution,
and opportunities for iterative patch improvement,
which can enhance patch quality.

\chatrep~\cite{xia2023keep} is a state-of-the-art conversational APR technique
that uses \chatgpt as the underlying LLM for bug repair.
It assumes a known buggy location and communicates with \chatgpt
for cloze-style or full-function patch generation
via a prompt that requests generating patched code
to replace the buggy location either as a line or a hunk (cloze-style)
or an entire method (full-function).
At the heart of \chatrep is its key component
performing iterative communication with \chatgpt for two purposes:
(1) fixing the previous failing patches generated by \chatgpt
in an attempt to obtain a plausible patch making all tests pass and
(2) generating alternative plausible patches
to improve patch diversity and increase the chance
of finding a correct patch.



While conversational APR techniques have exhibited superior repair abilities,
they still fail for a large number of real-world bugs,
even those having only one location to repair.
For example, according to a previous evaluation, 
\chatrep does not repair 175 (over 50\%) of 337 Defects4J bugs
whose developer patches (serving as the ground-truth)
change a line, a hunk (contiguous lines), or a method.
There is a lack of research investigating why conversational APR
fails for so many bugs, even the relatively simple ones.
The research is crucial, as it can provide critical guidance on
improving conversational LLM-based repair.


To bridge the gap, we took \chatrep as an exemplary conversational APR technique
and conducted a study to compare \chatrep's cloze-style and full-function
repair strategies, investigate the effectiveness of its key iterative component,
and analyze the failures. We implemented the \chatrep tool\footnote{By the time
we ran our experiments, the tool was unavailable.}
using ChatGPT (gpt-3.5-turbo) as the underlying conversational LLM
and two variants of \chatrep for comparison,
\chatreponeitersh and \chatreponeiterm,
which perform cloze-style and full-function repair with no iterative patch improvement.
We applied these tools to a sample of 53 Defects4J bugs
whose developer patches change a line, hunk, or method for repair.
We slightly adapted the prompts used by the tools,
requesting ChatGPT to provide not only a patch but also
an analysis of the problem and the program expected behavior.
We determined the correctness of a repair
by comparing the patch against the developer patch provided in the benchmark.
We also analyzed the repair failures based on the patch,
the description of the problem, and the program behavior given by ChatGPT.

The key findings of the study are as follows.
\begin{itemize}

    \item Cloze-style repair is prone to producing programs
    that do not compile. It is not as effective as
    the full-function repair, which simply asks ChatGPT to
    repair the whole method without indicating
    any buggy lines of the method.


    \item \chatrep's iterative approach for fixing previous failing patches 
    and finding alternative plausible patches does not appear to be helpful.
    Compared to \chatreponeiterm, which performs no iteration
    but independent patch generation using ChatGPT,
    \chatrep was not better and repaired even four fewer bugs.

    \item ChatGPT is not very good at repairing bugs
    whose fix ingredients used for patch construction
    are not native (e.g., not as operators or of language specific data types)
    and are located outside the buggy method.
    The success rate of repairing these bugs is 45\% 
    and is lower than those for bugs whose fix ingredients
    are all native (100\%) and are only within the scope of the buggy method (60\%).
        
    \item The main reasons for ChatGPT's failure are that
    (1) it misunderstands the failure and root cause;
    (2) it does not know the program expected behavior; and 
    (3) it fails to find the key fix ingredients for patch generation.
    
\end{itemize}

These findings suggest that method-level fault localization
is better suited for conversational LLM-based repair than
those targeting smaller code granularities such as statements;
that current iterative communication with ChatGPT does not fulfill
its potential in improving the patch quality;
and that future research on ChatGPT-based repair should focus on
helping ChatGPT understand the problem, infer the expected behavior,
and identify relevant fix ingredients.

\section{Research Questions} 
We seek to answer the following five research questions:
\begin{enumerate}
    \item \textbf{RQ1:} How effective are \chatreponeitersh and \chatreponeiterm?
    \item \textbf{RQ2:} How effective is \chatrep's iterative patch improvement?
    \item \textbf{RQ3:} Where are the fix ingredients of the failed repairs?
    \item \textbf{RQ4:} What is the connection between ChatGPT's analysis and the patch?
    \item \textbf{RQ5:} What are the reasons for ChatGPT's failures?
\end{enumerate}

For RQ1, we compare the effectiveness of cloze-style and full-function repair strategies.
For RQ2, we evaluate 
\chatrep's core iterative patch improvement component.
For RQ3, we analyze the locations of the fix ingredients used for patch construction
across different bugs and calculate the success rates of repair with them. 
For RQ4, we classify ChatGPT's analyses into fully correct, partially correct, and incorrect,
and then statistically evaluate the repair outcomes for each category. 
We do these to understand ChatGPT's responses and uncover the relationship 
between its analysis and the patch.
Finally, for RQ5, we identify and summarize the key reasons 
for ChatGPT's repair failures based on the issues explored in the previous research questions. 


\section{Experiment Setup, Method, and Results}
We implemented the \chatrep, \chatreponeitersh,
and \chatreponeiterm tools based on the algorithm and prompt examples 
provided in the paper~\cite{xia2023keep}.
As previously explained, we slightly modified the original prompts 
to assess the tools' abilities of problem understanding and expected behavior inference.
The code and all experimental results can be found
in our repository\footnote{\url{https://github.com/Aric3/an-implementation-of-chatrepair}},
in which the README file includes the implementation details of the tools.

We chose the widely used Defects4J dataset as a benchmark.
Because \chatrep currently only supports repairing a single function,
we only considered single-function bugs for experiments.
Due to the cost incurred by the manual patch review
and repair result assessment, we used a sample of the single-function bugs
in the original dataset.
To have the sample, we randomly selected
from each of the six projects (Lang, Chart, Closure, Mockito, Math, and Time) 
5 single-hunk (including single-line) bugs and at most
5 multi-hunk bugs that have two or three hunks to repair
(depending on the number of such bugs in the project). 
In this way, we obtained 30 single-hunk and 23 multi-hunk bugs.
Following previous evaluation of APR techniques~\cite{xia2023keep,xia_less_2022}, 
we determined the correctness of each patch manually
by comparing the patch against the developer patch
and checking whether they are semantically equivalent.

\subsection{RQ1: How effective are \chatreponeitersh and \chatreponeiterm?}

We selected 30 single-hunk bugs and ran both tools three times to repair each bug. 
This process yielded a total of 180 repair results, with 90 results 
generated by each tool.

\begin{table}[ht]
  \centering

  \caption{Comparison of \chatreponeitersh and \chatreponeitersh.
  RT: Repetition Times; CE: Percentage of Compilation Errors; CP: Percentage of Correct Patches.}
\label{tab:compare_oneiter_sh_m}
  \begin{threeparttable}
    \begin{tabular}{c|r|r|r}
      \hline
      \textbf{Method} & \textbf{RT} & \textbf{CE} & \textbf{CP} \\
      \hline
      \chatreponeitersh & 3  & 58.9\% & 6.7\% \\
      \chatreponeiterm & 3  & 11.1\% & 23.3\% \\
      \hline
    \end{tabular}
  \end{threeparttable}
\end{table}

TABLE~\ref{tab:compare_oneiter_sh_m}
shows that 58.9\% of the repairs given by \chatreponeitersh
have compilation errors, 
while \chatreponeiterm's repairs have much fewer (only 11.1\%) compilation errors.
\chatreponeiterm's correct repair rate is 23.3\%
whereas \chatreponeitersh's rate is only 6.7\%.
This shows that \chatreponeiterm is significantly more effective
than \chatreponeitersh. 

We identified three types of compilation errors from \chatreponeitersh's result:
(1) the patch contains redundant context for the target location;
(2) the patch is not made at the target location; and
(3) the patch introduces undefined items such as variables.
The first two types are dominant, accounting for 92.5\% of the errors.
For \chatreponeiterm, compilation errors arose for two reasons:
the introduction of undefined items and incomplete code generation.
%

\conclusion{Finding 1: The cloze-style repair strategy
generates a substantial number (58.9\%) of invalid patches
with compilation errors.}

\chatreponeitersh generated correct patches in 6 cases 
for 5 bugs, and \chatreponeiterm found correct patches in 21 cases for 9 bugs.
Lang-24 and Lang-51 are two bugs that were only repaired by \chatreponeitersh
but not by \chatreponeiterm. The buggy methods
for these two bugs are long, and they both exceed 50 lines,
making ChatGPT difficult to identify the exact location to repair.
The remaining 3 bugs correctly repaired 
by \chatreponeitersh were also repaired by the \chatreponeiterm.
 16 of the 21 correct patches made by \chatreponeiterm are significantly 
different from the developer patches in terms of the syntax.
In contrast, only 1 of the 6 correct patches made by \chatreponeitersh
are syntactically different from the developer patch. 

\conclusion{Finding 2: The full-function repair strategy
can generate a diverse range of patches, increasing the likelihood of
finding a correct patch. It however is not very effective at repairing
long methods.}

\subsection{RQ2: How effective is \chatrep's iterative patch improvement component?}

To evaluate \chatrep's iterative patch improvement,
we set the maximum number of attempts used by \chatrep to 24
(Xia et al. reported an average of 21.86 attempts used by \chatrep for plausible patch generation).
We also set the number of repetitions for \chatreponeiterm to 24. 
This setup ensures that both \chatrep and \chatreponeiterm 
call ChatGPT API 24 times per bug, enabling a fair comparison of
the iterative patch improvement component versus independent repeated prompting.
Our result shows that the \chatreponeiterm method repaired 23 bugs,
whereas the \chatrep method repaired only 18.
Moreover, during the plausible patch generation,
\chatrep produced duplicate patches that constituted 65\% of the total patches.

\conclusion{Finding 3: \chatrep's iterative patch improvement 
demonstrates no significant advantage over independent repeated prompting. 
Moreover, its plausible patch generation process produces 
a substantial proportion of duplicate patches (65\%), 
weakening its overall effectiveness.}

\subsection{RQ3: Where are the fix ingredients for the failed repairs?}

We referenced Yang et al.'s work~\cite{yang_where_2021} to
classify the fix ingredients into three distinct categories based on their sources.
The \textit{intrinsic} category refers to fix ingredients defined as
native tokens 
including operators and keywords such as basic data types and control structures. 
The \textit{local method} category encompasses fix ingredients 
retrieved from the buggy method. 
Since \chatrep operates 
at the method level, we group the remaining fix ingredient categories
under the label ``others''. This is because fix ingredients from these categories
do not appear in the provided prompts. We used scripts from Yang et al.'s code repository\footnote{https://github.com/DehengYang/repair-ingredients} to analyze 
and determine the fix ingredient category for each bug.

Across all repair experiments, 28 bugs were successfully repaired. 
We found that all five bugs requiring fix ingredients
at the \textit{intrinsic} level were successfully fixed, 
while the repair rate for bugs with fix ingredients 
at the \textit{local method} level is 60\%. For the remaining 38 bugs, which required fix ingredients 
beyond the \textit{local method} level, the repair rate dropped to 45\%. 
Although a correct patch does not need to exactly match the developer patch, the fix ingredient level provides valuable insight into the complexity of the code elements required for bug repair. 
When addressing bugs whose fix ingredient scopes
are \textit{intrinsic} and \textit{local method}, ChatGPT only needs to generate
built-in keywords, standard library functions, or ingredient defined within the buggy method,
which avoids the need to incorporate external project dependencies,
making these bugs comparatively easier to repair.

\conclusion{Finding 4: ChatGPT struggles with bug repair
that needs fix ingredients from outside the buggy method. 
The success rate of such repair is 45\%,
significantly lower than those of repairs requiring 
intrinsic (100\%) or local method (60\%) fix ingredients.}

\subsection{RQ4: What is the connection between ChatGPT's analysis and the patch?}
\label{sec:rq4}

The patch generation instructions provided at the end of the prompt
explicitly requested ChatGPT to provide an analysis of the issue
following a standardized format illustrated in an example. 
We collected and manually analyzed 249 repair results 
from \chatreponeiterm and \chatreponeitersh. 
To classify the analysis results, 
we used the following questions to establish three evaluation criteria.
\begin{enumerate}
\item Does the response identify the erroneous code lines? 
\item Does the response explain the reason for the error? 
\item Does the response clarify the logic behind the failure? 
\end{enumerate}

A fully correct analysis must satisfy all three criteria. 
Partially correct analyses are further categorized into three types, 
which are (1) Partial Explanation of the Reason: 
The response addresses only one or two of the outlined criteria; 
(2) Superficial Explanation: 
the response fails to meet any specific criterion and 
provides only a general description of the test case failure; 
and (3) Explanation with Extra Errors: The response 
includes correct explanations that satisfy the criteria, but it also 
contains additional incorrect explanations. 
If none of the criteria are met, the analysis is classified as incorrect.

\begin{table}[htbp]
  \centering
  \caption{The proportion of correct repairs under varying analysis categories. SH: single-hunk and single-line; MH: Multi-Hunk; CP, CA, PCA, and IA are the numbers of correct patches, correct analyses,
  partially correct analyses, and incorrect analyses, respectively.}
  \label{tab:TABLE2}
  \resizebox{0.8\linewidth}{!}{ 
    \begin{threeparttable}
      \begin{tabular}{c|c|r|r|r}
        \hline
        \textbf{Bug Type} & \textbf{Method} & \textbf{CP/CA}   & \textbf{CP/PCA} & \textbf{CP/IA} \\
        \hline
        SH   & \chatreponeitersh   & 10/21    & 1/8   & 0/61 \\
        \hline
        SH  & \chatreponeiterm   & 17/29    & 1/15 &  1/46 \\
        \hline
        MH   & \chatreponeiterm   & 8/16   & 1/30  &  0/23\\
        \hline
      \end{tabular}
    \end{threeparttable}
  }
\end{table}

Our statistical analysis reveals a strong correlation between ChatGPT's 
problem analysis and the quality of its patches. According to Table~\ref{tab:TABLE2},
when ChatGPT correctly understands the problem, 
the success rate for generating correct patches is 37.8\% (25/66). 
In contrast, the rate drops significantly when the analysis is incomplete
to less than 5.7\% (3/53) for partially correct analyses and 
to only 0.8\% (1/130) for incorrect analyses.

\conclusion{Finding 5: 
ChatGPT rarely makes correct patches if it does not understand the problem.}

\subsection{RQ5: What are the key reasons for ChatGPT's failures?}

We examined 210 instances of failed repairs from \chatreponeiterm and \chatreponeitersh and 
identified three reasons for ChatGPT's repair failures. 
The first reason is that ChatGPT fails to understand the root cause of the failure.
For 66\% of the repairs, ChatGPT gave incorrect problem analysis.
From the results of RQ4, when the problem analysis is incorrect, 
ChatGPT can rarely make a correct repair. Future research should 
investigate providing more effective code- and failure-related information
to help the LLM achieve better problem understanding.

The second reason is that ChatGPT does not know
the expected program behavior, that is, the behavior of the correct program.
This is especially the case when the repair
involves adding new code, as it is often difficult to infer a missing behavior.
Although the failing test case has assertions
that encode the expected execution outcome,
ChatGPT can still have difficulty understanding what is expected
as the final output or the internal state at the repair location.

The third reason is that ChatGPT fails to find the key fix ingredients
for patch generation. As discussed in Section~\ref{sec:rq4},
ChatGPT is not highly effective at using the fix ingredients
from the local method (success rate 60\%) for patch generation.
It can also often fail to generate a patch 
that needs fix ingredients beyond the local method scope (success rate 45\%),
which is understandable since the prompt used by
ChatGPT does not include fix ingredients from outside the buggy method.

In addition to these primary reasons,
others 
include failure to understand the logic of the buggy method,
failure to understand the logic of methods invoked by the buggy method, 
misunderstanding of prompt instructions, 
incomplete code generation, and failed test case overfitting.
Overall, 66\% of the failed repairs are related
to problem understanding, 
27\% to expected behavior inference, 
and 15\% to fix ingredient search.
The remaining reasons collectively accounted for less than 10\%. 
Note that the numbers do not add up to 100\%,
as a failed repair may be due to multiple reasons.

\conclusion{Finding 6: 
The main reasons for ChatGPT's repair failures are
(1) it fails to understand the root cause of the failure; 
(2) it does not know the expected program behavior;
and (3) it fails to find the key fix ingredients for patch generation.
}

\section{Related Work}

Sobania et al.~\cite{sobania2023analysis} evaluated 
ChatGPT's repair performance using a simple prompt
including only the buggy code and a query for repair without iteration.
They conducted experiments on the QuixBugs dataset
containing only 40 programming-level bugs.
Their results may not reflect the LLM's ability in repairing more complex real-world bugs.
Zhang et al.~\cite{zhang_critical_2023} developed an iterative repair approach
based on ChatGPT and evaluated its performance on the EVALGPTFIX dataset
containing competition bugs.
They classified the competition bugs into three categories
and designed prompts for each. The prompts and the approach may not be 
suitable for real-world bug repair.
Xia et al.~\cite{xia2023keep} introduced \chatrep, 
a state-of-the-art ChatGPT-based conversational repair method.
Their work lacks an analysis of the responses generated by ChatGPT 
that include both textual explanations and code patches.
In our work, we evaluated the repair effectiveness of \chatrep
on Defects4J, focusing particularly on bugs it failed to repair.
Through a detailed analysis, we identified and summarized the reasons
behind these repair failures from multiple perspectives.

\section{Threats to Validity}
Potential errors from tool implementation and
manual analysis pose threats to the validity of the study,
although we have carefully tested the tools
and checked our results. We released the tool and results
for public review and examination. 
Our results may not generalize to other LLM models or datasets, 
and they could be influenced by the randomness of ChatGPT.
We hope to conduct a larger-scale study as future work.



\section{Conclusion and Future Work}

We conducted a study to understand the effectiveness
and failures of conversational LLM-based APR.
The study is based on the state-of-the-art technique \chatrep.
We investigated \chatrep's cloze-style and full-function
repair strategies, assessed its core iterative component
for patch improvement, and analyzed its repair failures.
The study has led to several findings
that we believe provide important implications for APR research.
We are currently exploring approaches to improving ChatGPT's 
understanding of problems for better patch generation.
Future work includes conducting extensive experiments
that incorporate more conversational APR techniques 
and evaluate them on diverse benchmarks,
particularly those not subject to data leakage.

\section{Acknowledgement}

This work was partially supported by the National Natural Science Foundation of China
under the grant numbers 62202344 and 62141221 and the OPPO Research Fund.

\bibliographystyle{IEEEtran}
\bibliography{paper}

\begin{thebibliography}{1}
\providecommand{\url}[1]{#1}
\csname url@samestyle\endcsname
\providecommand{\newblock}{\relax}
\providecommand{\bibinfo}[2]{#2}
\providecommand{\BIBentrySTDinterwordspacing}{\spaceskip=0pt\relax}
\providecommand{\BIBentryALTinterwordstretchfactor}{4}
\providecommand{\BIBentryALTinterwordspacing}{\spaceskip=\fontdimen2\font plus
\BIBentryALTinterwordstretchfactor\fontdimen3\font minus \fontdimen4\font\relax}
\providecommand{\BIBforeignlanguage}[2]{{%
\expandafter\ifx\csname l@#1\endcsname\relax
\typeout{** WARNING: IEEEtran.bst: No hyphenation pattern has been}%
\typeout{** loaded for the language `#1'. Using the pattern for}%
\typeout{** the default language instead.}%
\else
\language=\csname l@#1\endcsname
\fi
#2}}
\providecommand{\BIBdecl}{\relax}
\BIBdecl

\bibitem{xia2023keep}
C.~S. Xia and L.~Zhang, ``Keep the conversation going: Fixing 162 out of 337 bugs for \$0.42 each using {ChatGPT},'' \emph{arXiv preprint arXiv:2304.00385}, 2023.

\bibitem{xia_less_2022}
{C. S. Xia and L. Zhang}, ``\BIBforeignlanguage{en-US}{Less training, more repairing please: revisiting automated program repair via zero-shot learning},'' in \emph{\BIBforeignlanguage{en-US}{ESEC/FSE}}, 2022, pp. 959--971.

\bibitem{yang_where_2021}
D.~Yang, K.~Liu, D.~Kim, A.~Koyuncu, K.~Kim, H.~Tian, Y.~Lei, X.~Mao, J.~Klein, and T.~F. Bissyandé, ``\BIBforeignlanguage{en}{Where were the repair ingredients for {Defects4J} bugs?}'' \emph{\BIBforeignlanguage{en}{Empir Software Eng}}, vol.~26, no.~6, p. 122, 2021.

\bibitem{sobania2023analysis}
D.~Sobania, M.~Briesch, C.~Hanna, and J.~Petke, ``An analysis of the automatic bug fixing performance of {ChatGPT},'' \emph{arXiv preprint arXiv:2301.08653}, 2023.

\bibitem{zhang_critical_2023}
Q.~Zhang, T.~Zhang, J.~Zhai, C.~Fang, B.~Yu, W.~Sun, and Z.~Chen, ``A critical review of large language model on software engineering: {An} example from {ChatGPT} and automated program repair,'' \emph{arXiv preprint arXiv:2310.08879}, 2023.

\end{thebibliography}

\end{document}